\documentclass[11pt,a4paper]{article}
\usepackage{jcappub}

\usepackage{epsfig,psfrag,graphics,verbatim}
\usepackage{graphicx}

\usepackage{dcolumn}
\usepackage{bm}
\usepackage{epsfig}
\usepackage{amsmath,amssymb}
\usepackage{psfig}
\usepackage{ulem}
\usepackage{natbib}

\def \apj  {ApJ}

\def \apjl  {ApJL}
\def \prd {Phy.Rev.D}
\def \mnras {MNRAS}
\def \chisq  {\ifmmode  \chi^2   \else  $\chi^2$  \fi}  
\def \spose#1{\hbox  to 0pt{#1\hss}}  
\def \lta{\mathrel{\spose{\lower 3pt\hbox{$\sim$}}\raise  2.0pt\hbox{$<$}}}
\def \gta{\mathrel{\spose{\lower  3pt\hbox{$\sim$}}\raise 2.0pt\hbox{$>$}}}

\title{The Galactic Halo in Mixed Dark Matter Cosmologies}

\author[a]{D. Anderhalden}
\author[a]{J. Diemand}
\author[b]{G. Bertone}
\author[c]{\\A. V. Macci\`o}
\author[a]{A. Schneider}

\affiliation[a]{Institute for Theoretical Physics, University of Z\"urich, Winterthurerst. 190, 8057 Z\"urich, Switzerland}
\affiliation[b]{GRAPPA Institute, University of Amsterdam, Science Park 904, 1090 GL Amsterdam, Netherlands}
\affiliation[c]{Max-Planck-Insitute for Astronomy, K\"onigstuhl 17, 69117 Heidelberg, Germany}
\emailAdd{donninoa@physik.uzh.ch}
\emailAdd{diemand@physik.uzh.ch}
\emailAdd{gf.bertone@gmail.com}
\emailAdd{maccio@mpia.de}
\emailAdd{aurel@physik.uzh.ch}
\abstract{
A possible solution to the small scale problems of the cold dark matter (CDM) scenario is that the dark matter consists of two components, a cold and a warm one. We perform a set of high resolution simulations of the Milky Way halo  varying the mass of the WDM particle ($m_{\rm WDM}$) and the cosmic dark matter mass fraction in the WDM component ($\bar{f}_{\rm W}$).  The scaling ansatz introduced in combined analysis of LHC and astroparticle searches postulates that the relative contribution of each dark matter component is the same locally as on average in the Universe (e.g. $f_{\rm W,\odot} = \bar{f}_{\rm W}$). Here we find however, that the normalised local WDM fraction ($f_{\rm W,\odot}$ / $\bar{f}_{\rm W}$) depends strongly on $m_{\rm WDM}$ for $m_{\rm WDM} <$ 1 keV. Using the scaling ansatz can therefore introduce significant errors into the interpretation of dark matter searches. To correct this issue a simple formula that fits the local dark matter densities of each component is provided.
}


\begin{document}
\maketitle

\section{Introduction}
\label{introduction}

Structure formation in the Universe is dominated by a mysterious dark matter component, whose energy density is about six times higher than the energy density made up of Standard Model particles. The kinds of structures which may form and their internal properties depends strongly on the as yet unknown physical nature of the dark matter particle. A common assumption is that dark matter is made of heavy, cold thermal relic particles that decoupled from normal matter in the very early Universe, generically referred to as cold dark matter (CDM) \cite{Peebles1982,Blumenthaletal1984}.

Whilst there is a large body of indirect astrophysical evidence that supports CDM, especially on large scales, there are significant indications of possible shortcomings on small scales. First of all, CDM galaxy haloes contain a huge number of subhaloes \cite{Mooreetal1999c,Klypinetal1999,Diemandetal2008,Springeletal2008,Stadeletal2009}, while observations indicate that only relatively few satellite galaxies exist around the Milky Way and M31 \cite{Mateo1998,Koposovetal2008}. One possible way out of this problem would be to have no or only very inefficient star formation in smaller CDM haloes \cite{Kravtsov2003,Maccioetal2010}. However, such models populate all the larger subhaloes with dwarf galaxies.  Furthermore, the central densities of these largest CDM subhaloes seem to be about two times higher than the densities inferred form the observed stellar kinematics of the dwarf satellite galaxies \cite{Boylan-Kolchinetal2011,Lovelletal2011,Rashkovetal2011}.\newline
\indent Secondly, the inner density profiles inferred from galaxy rotation
curves seem to be flatter than the NFW profile \cite{Navarroetal1997} used to fit
simulated CDM halos (eg. \cite{Simonetal2005,Gentileetal2009}). At the
relevant scale of one percent of the virial radius the typical
logarithmic density slopes are significantly steeper than the NFW form
and there is substantial halo to halo scatter $\gamma = 1.26 \pm 0.17$
(e.g. \cite{Diemandetal2004}). Fitting functions which are steeper than NFW at that scale
provide better fits (e.g. \cite{Navarroetal2004,Merritt2006,Diemandetal2004,Diemandetal2008,Springeletal2008,Stadeletal2009}, which further
increases the tension with the observations.


Thirdly the observed number of dwarf galaxies in voids is smaller than expected from CDM \cite{PeeblesNusser2010}.

One possible solution to these problems may be that dark matter is not cold, but warm (WDM). In this scenario dark matter particles are created relativistic unlike 
its CDM counterpart and so `free-stream' until relatively late times. Following \cite{KolbTurner}, the free streaming length in the non-relativistic regime can be approximated by
\begin{equation}
\lambda_{FS} \simeq r_H(t_{NR}) \Big[1 + \frac{1}{2}\log\frac{t_{EQ}}{t_{NR}} \Big],
\end{equation}
where $t_{EQ}$ denotes the epoch of matter-radiation equality after which the Jeans length starts to decrease. $t_{NR}$ is the time when the dark matter particles becomes non-relativistic. Due to the large mass of a cold thermal relic and its very early freeze out, the comoving horizon, $r_H$, is extremely small compared with its warm dark matter counterpart. $t_{NR}$ strongly depends on the mass of the particle, that in turn therefore determines the characteristic mass scale below which primordial fluctuations are erased (e.g. \cite{Schneider2011}).

There are several models motivated from particle physics that provide potential WDM candidates. The dynamics of the dark matter particle is mainly determined by its production history, i.e. if it once was in thermal equilibrium or not. The two most prominent candidates for these scenarios are the thermally produced {\it gravitino} \cite{Ellisetal1984} and the non-thermally produced {\it sterile neutrino} \cite{DodelsonWidrow1994,Shaposhnikov2007,boyarsky2009}. 

Refs. \cite{Dunkley2011} and \cite{Keisler2011} recently reported a two sigma preference for an increased damping tail of the CMB power spectrum. As a possible explanation, they claim strong evidence for a fourth relativistic species. This is in line with measurements from atmospheric \cite{Aguilar2010} and nuclear reactor \cite{Mention2011} experiments, which find some evidence in favour of a sterile neutrino. Such light sterile neutrinos could be part of the cold + hot (rather than cold + warm) dark matter models. \cite{Polisensky2011} have listed approximate conversion formulas for the mass of the warm particle for the most common production mechanisms, i.e. the thermal production of the gravitino \cite{Vieletal2005}, the resonant \cite{ShiFuller1999} and non-resonant production \cite{DodelsonWidrow1994} of the sterile neutrino via oscillations with active neutrinos, and the sterile neutrino production via the decay of a scalar field \cite{Shaposhnikov2006,Kusenko2006}:
\begin{eqnarray}
&&m_{\nu_s}^{DW} \simeq 4.379 \text{ keV } \Big(\frac{m_{\text{WDM}}}{1 \text{keV}} \Big)^{4/3} \nonumber \\
&&m_{\nu_s}^{DW} / m_{\nu_s}^{SF} \simeq 1.5  \\
&&m_{\nu_s}^{DW} / m_{\nu_s}^{scalar} \simeq 4.5. \nonumber
\end{eqnarray}
The power spectrum suppression of the latter two models is different from the usual thermally produced WDM and the above formulae give therefore only a rough estimate. These WDM candidates might help alleviate the small-scale issues of CDM, while maintaining the same large scale structure. However, the core created in such WDM models might not be large enough to match the observed inner density profiles of dwarf galaxies (e.g. \cite{maccio2012,villa11}, and references therein).

Another viable solution would be to have a mixture of CDM plus WDM -- the C+WDM or mixed dark matter (MDM) model (e.g. \cite{Boyarskyetal2009a,Boyarskyetal2009b,boyarsky2009}). Various theoretical motivations exist for C+WDM. The sterile neutrino for instance can be created via different production mechanism and appear in two disjunct, effective states, a cold and warm one (see \cite{Boyarskyetal2009a} for details). One could also think of, besides the 'usual' cold neutralino candidate, to have a second dark matter particle, e.g. a thermal gravitino. 

Recently, a lot of effort has been dedicated to resolve the small-scale structure of CDM haloes \cite{Diemandetal2008,Springeletal2008,Stadeletal2009}, however only a few simulations have been performed to study the WDM scenario \cite{colin2008,Polisensky2011,Lovelletal2011}, and a single one for the C+WDM model \cite{MaccioMixed}. The aim of this work is to determine the galactic dark matter halo properties in mixed dark matter cosmologies, especially the local dark matter densities of each component. A detailed analysis of substructure abundance and kinematics will be presented in a second paper (Anderhalden et al. 2012b, in prep.).

This paper is organised as follows. In Section \ref{simulations} the computation of the mixed dark matter transfer function is shown, as well as a summary of the numerical simulations. The internal structure of the Milky Way halo, i.e. density and phase space density profiles as well as the local dark matter abundance, is presented in Section \ref{internalstructure}. Implications of our results on dark matter detection are described in Section \ref{detection}.
\section{Numerical Methods}
\label{simulations}

The transfer function of any non-standard cosmology can be written as the square root of the ratio of the power spectrum of cosmology $i$ to that of CDM, i.e.
\begin{equation}
T_i(k) = \Big(\frac{P_i(k)}{P(k)_{\text{CDM}}} \Big)^{1/2}.
\end{equation}
The transfer function in a pure, thermally produced warm dark matter scenario is well fitted by \cite{Bodeetal2001,Vieletal2005}
\begin{equation}\label{wdmfit}
T_{\text{WDM}}(k) = \Big(1 + \big(\alpha k \big)^{2\nu} \Big)^{-5/\nu},
\end{equation}
where $\nu=1.12$ and $\alpha$ is a function of the warm particles mass,
\begin{equation}
\alpha = 0.049 \Big(\frac{m_{x}}{\text{keV}} \Big)^{-1.11} \Big(\frac{\Omega_x}{0.25} \Big)^{0.11} \Big(\frac{h}{0.7} \Big)^{1.22} h^{-1} \text{Mpc}.
\end{equation}
The above form of the transfer function leads to a steep cutoff in the linear power spectrum, resulting in a strong suppression of small structures below the free-streaming mass (for a detailed discussion see e.g. \cite{Vieletal2005,Boyarskyetal2009a}).

The qualitative behavior of the mixed C+WDM power spectrum differs from the pure warm dark matter case in one important aspect. Although the mass of the WDM particle still determines the cutoff scale, the presence of a cold component stabilises the drop-off and yields the transfer function to approach a constant plateau \cite{Boyarskyetal2009a}. The size of this step is to first order completely described by the fraction of warm dark matter, i.e.
\begin{equation}\label{wdmfraction}
\bar{f}_{\rm W} = \frac{\Omega_{\text{WDM}}}{\Omega_{\text{WDM}}+\Omega_{\text{CDM}}} = \frac{\Omega_{\text{WDM}}}{\Omega_{\text{DM}}}.
\end{equation}

\begin{figure}
\centering
\includegraphics[scale=0.5]{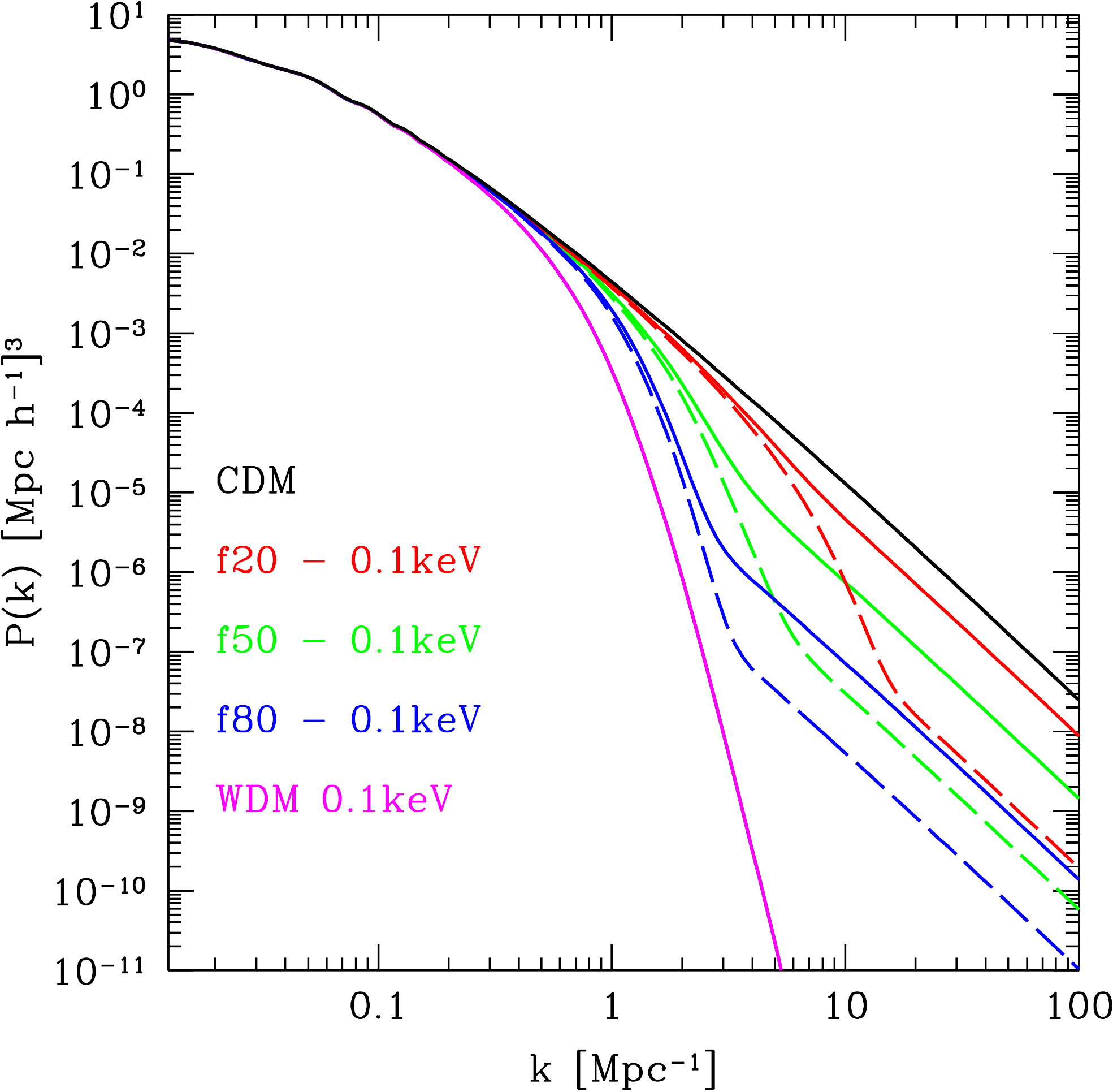}
\caption{Evolution of the linear mixed dark matter power spectrum when keeping the mass of the warm particle fixed and varying the fraction $\bar{f}_{\rm W}$. For reference the $\bar{f}_{\rm W}=0$ (pure CDM) and $\bar{f}_{\rm W}=1$ (pure WDM) cases are plotted too (black solid lines). In each case the colored solid lines refer to the cold component, whereas the colored dashed lines refer to the corresponding warm component of the mixed dark matter fluid.}
\label{linpower}
\end{figure}

Unlike Ref. \cite{Boyarskyetal2009a}, we compute the transfer function of the cold and the warm component separately (even if they are not independent from one another, see Section \ref{mdmtransferfunction}). To illustrate this, a set of C+WDM linear power spectra at redshift $z=99$ are plotted in Fig. \ref{linpower}. By keeping the mass of the warm particle constant (in this particular case $m_{\text{WDM}} = 0.1$keV) and varying the fraction $\bar{f}_{\rm W}$ from zero (CDM) to one (WDM), one can see that the plateau, i.e. the step size, is dictated by the fraction, whereas the mass of the warm dark matter particle is responsible for the cutoff, i.e. the scale where the transfer function starts to differ from unity. Solid lines refer to the cold component of the mixed fluid, dashed lines to warm one. The higher the amount of warm dark matter, the more the plateaus of the two components approach each other, until finally a pure warm dark matter cosmology is reproduced ($\bar{f}_{\rm W} = 1$). In the following we shortly describe how the aforementioned transfer functions are computed.

\subsection{Mixed Dark Matter Transfer Function}\label{mdmtransferfunction}
In order to run N-body simulations containing both cold and warm dark matter particles in a distinguishable way, the transfer function of each component has to be specified individually. On large scales, dark matter is very well described by a perfect fluid approximation. To study the behaviour of small perturbations embedded in a homogeneous and isotropic background, it is appropriate to linearize the complete set of equations (i.e. Continuity, Euler and Poisson equation (e.g. \cite{Padmanabhan})). Since relativistic effects are negligible on scales below the horizon, we will restrict our computation to the Newtonian approximation at leading order, instead of solving the fully relativistic Boltzmann equation. 

The total dark matter density of the mixed (i.e. cold plus warm) fluid can be written as
\begin{equation}
\rho(\textbf{x}) = \rho_{\text{C}}(\textbf{x}) + \rho_{\text{W}}(\textbf{x}),
\end{equation}
where $\rho_{\text{c}}(\textbf{x})$ and $\rho_{\text{w}}(\textbf{x})$ refer to densities of the cold and warm component respectively. Using the dimensionless density parameter, $\delta \equiv \frac{\rho - \bar{\rho}}{\bar{\rho}}$, and defining $\bar{f}_{\rm C} \equiv 1 - \bar{f}_{\rm W}$ it follows
\begin{equation}
\delta_{\text{MDM}} = \bar{f}_{\rm W} \; \delta_\text{W} + \bar{f}_{\rm C} \; \delta_\text{C}.
\end{equation}
By taking the ensamble-average over all $k$-modes of $\delta$, $\langle | \delta_k |^2\rangle$, the power spectrum of the mixed dark matter fluid can be entirely expressed in terms of $\bar{f}_{\rm W}$ /  $\bar{f}_{\rm C}$ and the CDM power spectrum\footnote{We again want to emphasise that this is only true in linear perturbation theory, neglecting all higher order interaction terms.},
\begin{eqnarray}\label{mdmpower1}
P_{\text{MDM}}(k) &=& \bar{f}_{\rm W}^2 P_{\text{WDM}}(k)
		+ \; \bar{f}_{\rm C}^2 P_{\text{CDM}}(k) \nonumber \\
		&&+ \; 2 \bar{f}_{\rm W} \bar{f}_{\rm C} P_{\text{C,W}}(k) \nonumber \\
		&=& \Big[\bar{f}_{\rm W}^2 r_\text{W}^2 + \bar{f}_{\rm C}^2 r_\text{C}^2 + 2 \bar{f}_{\rm W} \bar{f}_{\rm C} r_\text{W} r_\text{C} \Big] P_{\text{CDM}}(k) \nonumber \\
		&=& T^2_{\text{MDM}}(k) P_{\text{CDM}}(k),
\end{eqnarray}
where we have introduced the ratios
\begin{equation}\label{ratios}
r_\text{C}(k) \equiv \frac{\delta_\text{C}(k)}{\delta_{\text{CDM}}} \,\, \text{and} \,\, r_\text{W}(k) \equiv \frac{\delta_\text{W}(k)}{\delta_{\text{CDM}}}.
\end{equation} 

In order to set up initial conditions with a power spectrum for both the cold and the warm component individually, we also have to account for the interference term in Eq. \eqref{mdmpower1}. This can be achieved by solving the linearized Newtonian perturbation equations for a dark matter fluid consisting of a cold and a warm dark matter particle\footnote{For simplification we assume that the baryonic density perturbations follow those of cold dark matter.} (e.g. \citep{Padmanabhan}):
\begin{eqnarray}\label{perturb1}
\ddot{\delta}_\text{C} + 2 \frac{\dot{a}}{a} \dot{\delta}_{\text{C}}& = & \frac{4 \pi G \rho_{0}}{a^3} \Big [\bar{f}_{\rm W}\;\delta_{\text{W}} + \bar{f}_{\rm C}\;\delta_\text{C}\Big ]  \\
\ddot{\delta}_\text{W} + 2 \frac{\dot{a}}{a} \dot{\delta}_{\text{W}}& = & \frac{4 \pi G \rho_{0}}{a^3} \Big [\bar{f}_{\rm W}\;\delta_{\text{W}} + \bar{f}_{\rm C}\;\delta_\text{C}\Big ] - \frac{k^2 \sigma_0^2}{a^4}\delta_{\text{W}} \nonumber,
\end{eqnarray}
where the derivatives are with respect to cosmic time and $\sigma_0$ is the rms velocity dispersion of the warm component. Following \cite{Bodeetal2001}, the velocity dispersion of the WDM particle at redshift $z$ can be expressed as
\begin{equation}\label{velocitydispersion}
\frac{v_0(z)}{1+z} = .012 \Big(\frac{\Omega_{\text{WDM}}}{0.3} \Big)^{\frac{1}{3}} \Big(\frac{h}{0.65} \Big)^{\frac{2}{3}}\Big(\frac{1.5}{g_X} \Big)^{\frac{1}{3}} \Big(\frac{\text{keV}}{m_{\text{WDM}}} \Big)^{\frac{4}{3}}\text{km}\;\text{s}^{-1},
\end{equation}
and the rms velocity is $\sigma_0 = 3.571 v_0$. $g_X$ in Eq. \eqref{velocitydispersion} is the number of degrees of freedom and can, for a generic thermal candidate, be set equal 1.5 \cite{Bodeetal2001}.
Rewriting Eqs. \eqref{perturb1} as a function of the scale factor, 
\begin{eqnarray}\label{perturb2}
\delta_\text{C}^{\prime\prime} + \frac{3}{2 a}\delta_\text{C}^{\prime} &=& \frac{9 \pi G \rho_0}{a^2}\Big[\bar{f}_{\rm W}\;\delta_\text{W} + \bar{f}_{\rm C}\;\delta_\text{C} \Big] \\
\delta_\text{W}^{\prime\prime} + \frac{3}{2 a}\delta_\text{W}^{\prime} &=& \frac{9 \pi G \rho_0}{a^2}\Big[\bar{f}_{\rm W}\;\delta_\text{W} + \bar{f}_{\rm C}\;\delta_\text{C} \Big] - \frac{9}{4}\frac{k^2 \sigma_0^2}{a^3}\delta_\text{w}, \nonumber
\end{eqnarray}
a direct way to solve the above set of coupled differential equations is provided. The derivatives are now with respect to the scale factor. In the Newtonian approximation, dark matter perturbations do not have a significant growing solution before matter-radiation equality. We therefore choose $a_{\rm eq}$ as the initial epoch for solving Eqs. \eqref{perturb2} numerically.
\begin{figure}
\centering
\includegraphics[scale=0.5]{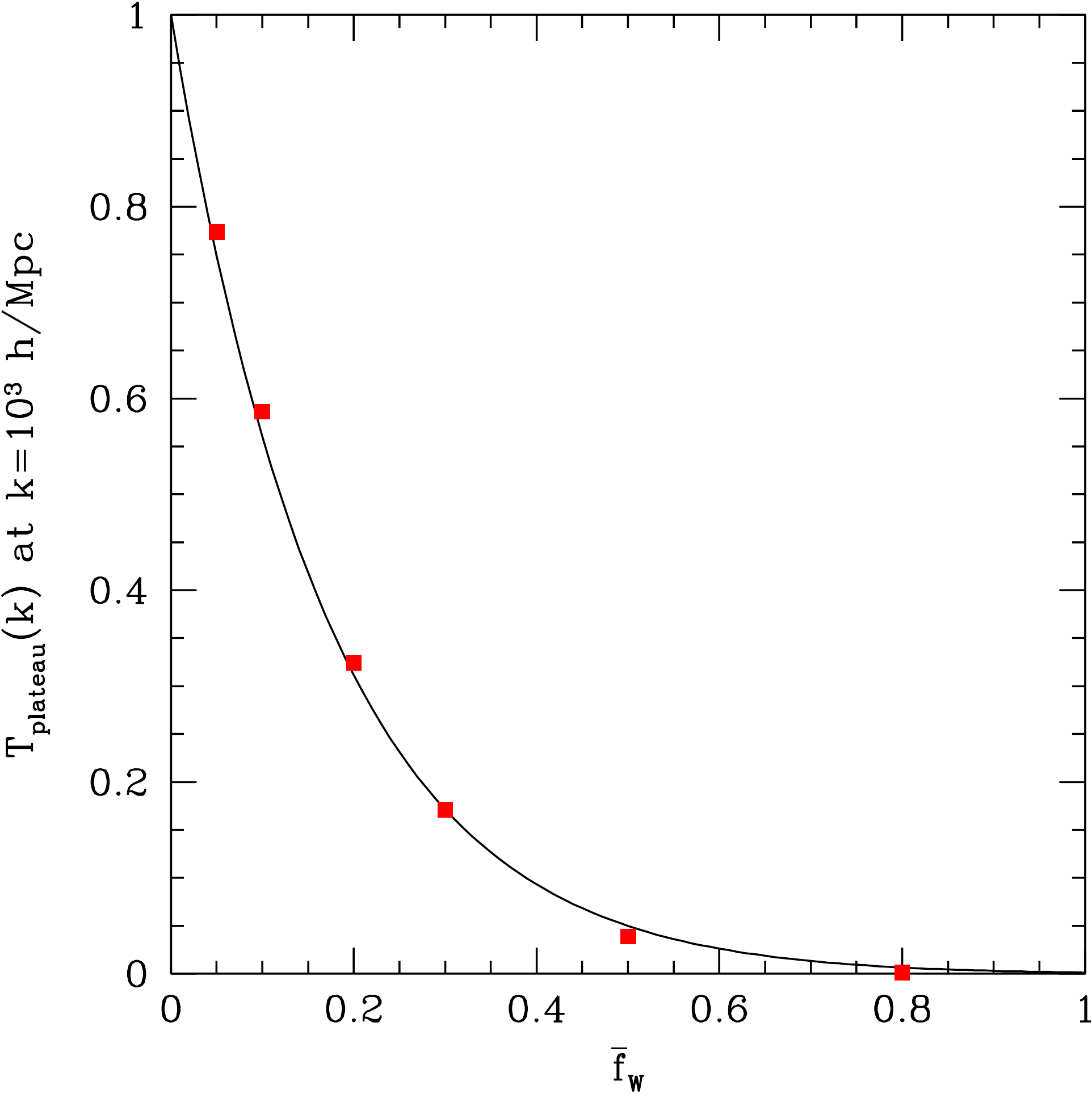}
\caption{Comparison between an empirical fitting function (solid black line, see Eq. \eqref{plateaufit}) for the plateau of the mixed dark matter transfer function and the one based on our two fluid approach (red squares).}
\label{comparison}
\end{figure}
For adiabatic fluctuations, linear theory allows us to calculate the density contrast $\delta_k(a_{\rm eq})$ to some time $a$ (in our case the starting redshift of the simulation), via a transfer function $T(k)$,
\begin{equation}
\delta(k,a_{\rm IC}) \propto T(k) \delta(k,a_{\rm eq}).
\end{equation}
It is straightforward now to calculate transfer functions for the cold and the warm component individually in such a way, that the power the of CDM model can be transformed into any mixed dark matter scenario:
\begin{equation}\label{mdmpower}
P_{\text{MDM}}^{\text{C,W}}(k) = P_{\text{CDM}}(k) \cdot T^2_{\text{C,W}}(k) \simeq P_{\text{CDM}}(k) \cdot r^2_{\text{C,W}}.
\end{equation}
The above equation now provides an easy method to setup initial conditions for cosmological simulations. To test the validity of our simplistic approach, we make use of the fact that the mixed dark matter transfer function reaches a constant plateau at high values of $k$, fully determined by the fraction $\bar{f}_{\rm W}$ and compute $T_{\text{plateau}}$ at $k=10^3$ h/Mpc for a set of different values of the warm dark matter fraction. Fig. \ref{comparison} shows the comparison between our results (red squares) and an empirical fitting function of \cite{Boyarskyetal2009a} (black solid line), describing a full Boltzmann code calculation for a mixed dark matter fluid. It is given by
\begin{equation}\label{plateaufit}
T_{\text{plateau}} = (1-\bar{f}_{\rm W})\cdot \Big( \frac{\Omega_m}{\Omega_r} g(a_0) \Big)^{-(3/4)\bar{f}_{\rm W}},
\end{equation}
where $g(a_0)$ is a function of the energy density content of the universe (see \cite{Boyarskyetal2009a} for details). It is evident that for small amounts of WDM the linearized approach slightly overestimates the exact value of $T_{\text{plateau}}$, whereas it slightly underestimates $T_{\text{plateau}}$ when the warm component is dominant. However, this effect is extremely small, so it is sufficient to set up initial conditions for numerical simulations based on the procedure described above.

\begin{figure}
\centering
\includegraphics[scale=0.5]{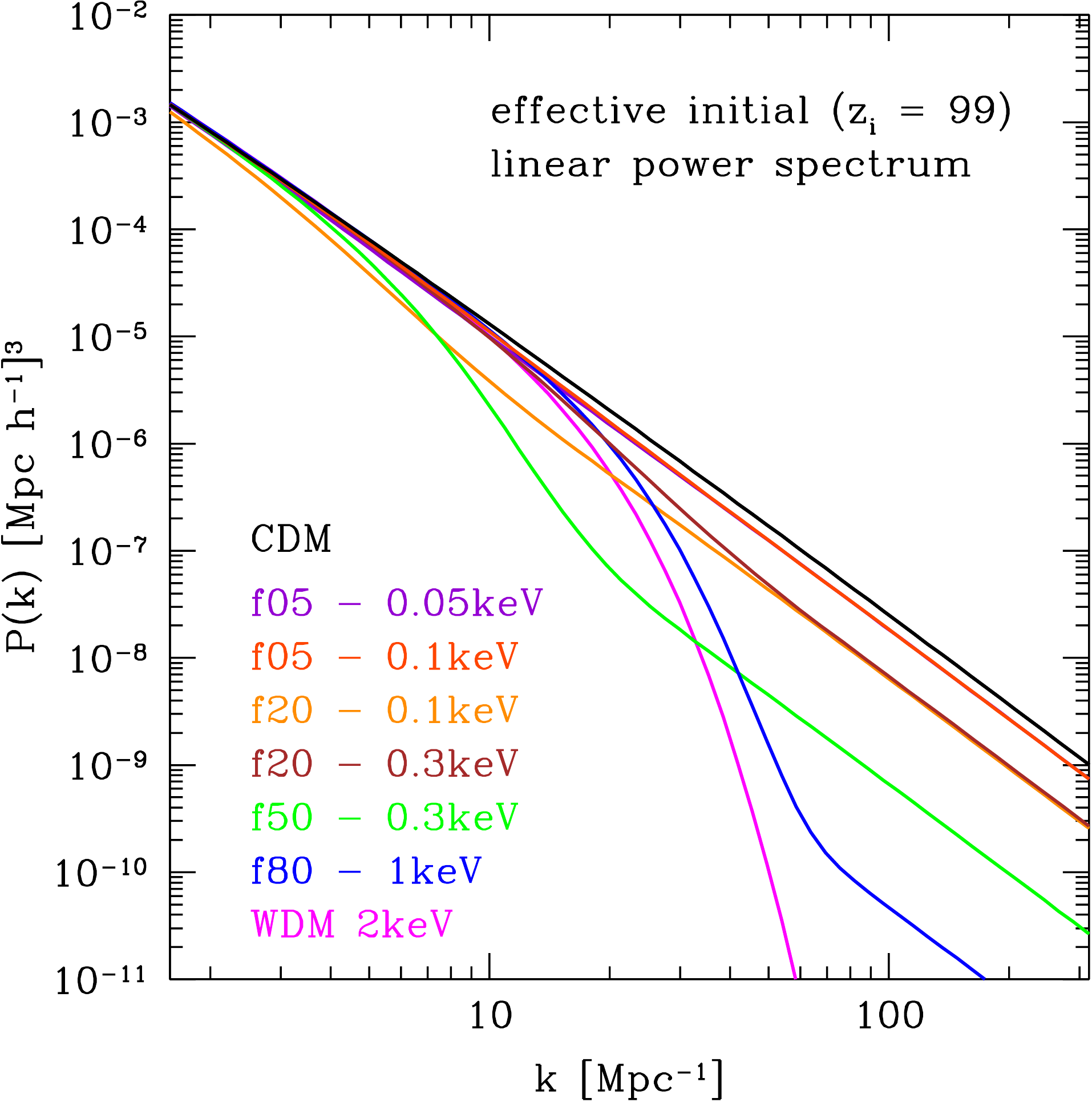}
\caption{Effective initial linear power spectra of all mixed dark matter fluids.}
\label{powermdmsimulation}
\end{figure}

Fig. \ref{powermdmsimulation} shows the effective initial power spectrum, calculated via Eq. \eqref{mdmpower1}, for all different cosmologies used in this work (see Table \ref{sims} for details). It is important to note that Fig. \ref{powermdmsimulation} shows the power spectra for the mixed, i.e. cold plus warm fluid. Our initial conditions are set up by using the individual transfer function for each component separately (illustrated in Fig. \ref{linpower}).

\subsection{Simulations}
All numerical simulations have been carried out using the treecode {\sc pkdgrav} \cite{Stadelpkd}. We perform six different mixed dark matter cosmologies as well as a pure CDM and a pure WDM simulation with a particle mass of $m_{\text{wdm}} = 2$keV. Our initial conditions are based on the WMAP7 cosmological model \cite{Komatsu2011}: $\sigma_8 = 0.8$, $h=0.7$, $\Omega_{\text{dm}}=0.227$, $\Omega_{\text{b}}=0.046$, $\Omega_{\Lambda}=0.727$, $n_\text{s}=0.961$ and are created with a parallel version of the GRAFIC package \cite{Bertschinger2001}. 

To begin with we run large scale simulations of two merged cosmological boxes with $L_{\text{box}} = 40$Mpc, all together using $2 \times 256^3$ particles. The merging of two boxes represents the mixture of the cold and warm component\footnote{We also choose that approach for the pure CDM \& WDM simulations in order to have a one to one correspondence in the final halo.} and is accomplished by offsetting one of the boxes by half the mean interparticle separation in each direction, to ensure that particles do not lie on top of each other. The initial conditions for all eight simulations have been created using identical random seeds. We then select an isolated, Milky Way sized halo of mass $M\sim10^{12}$M$_{\odot}$ and re-run it at $8^3$ higher mass resolution, leading to a dark matter particle mass of $m_p=1.38 \times 10^5$ M$_{\odot}$ and a physical gravitational softening of 355 pc in the refined region. The high resolution run was performed by a single level of spatial refinement, so in order to avoid mixing of heavy particles in the final halo, the Lagrangian region was chosen to be $3\times R_{200}$ of the original selected object. The total number of high resolution particles within $R_{200}$ of the refined halo depends on the cosmological model, but is of order $10^7$. Details of the simulations are listed in Table \ref{sims}.
\begin{figure*}
\includegraphics[scale=0.18]{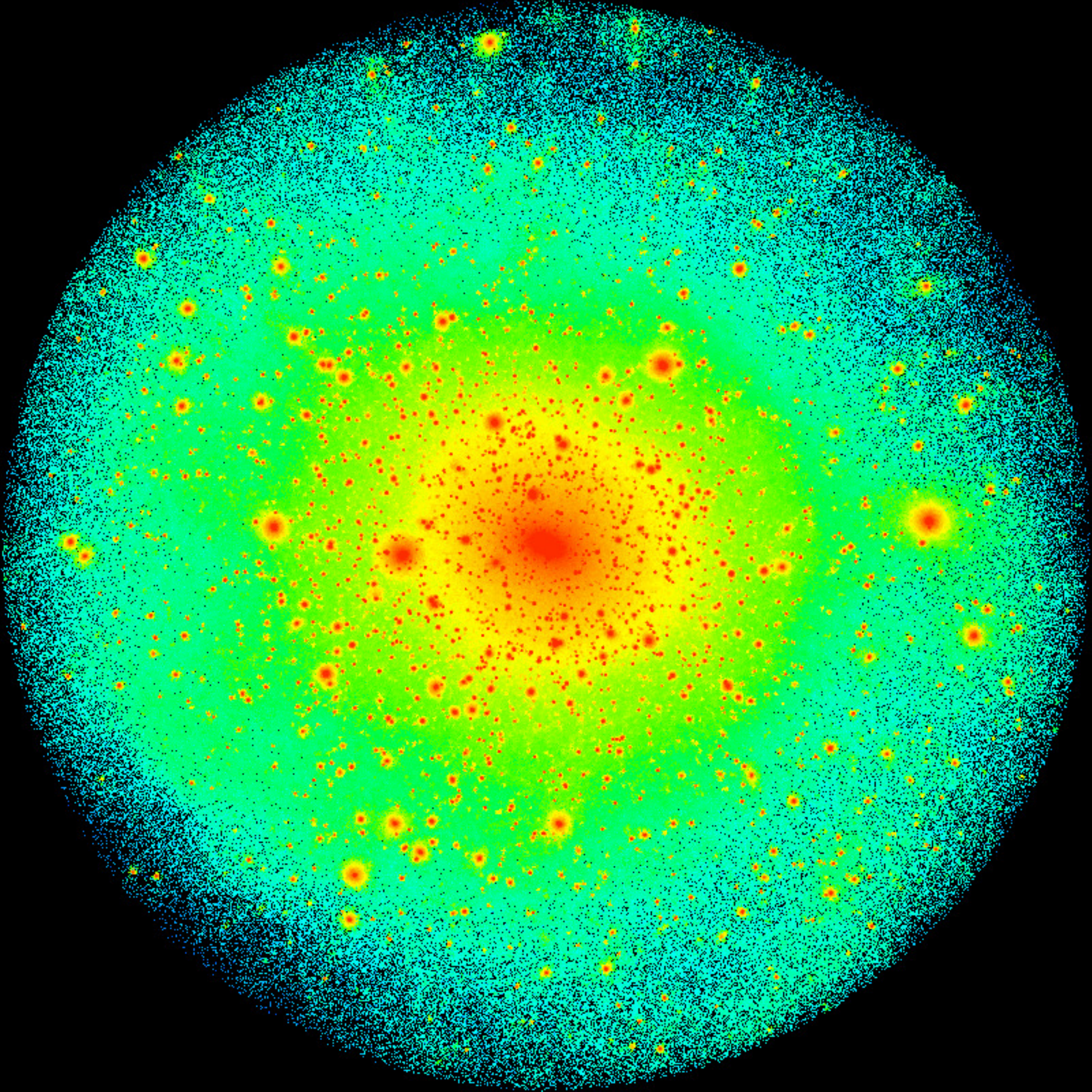}
\includegraphics[scale=0.18]{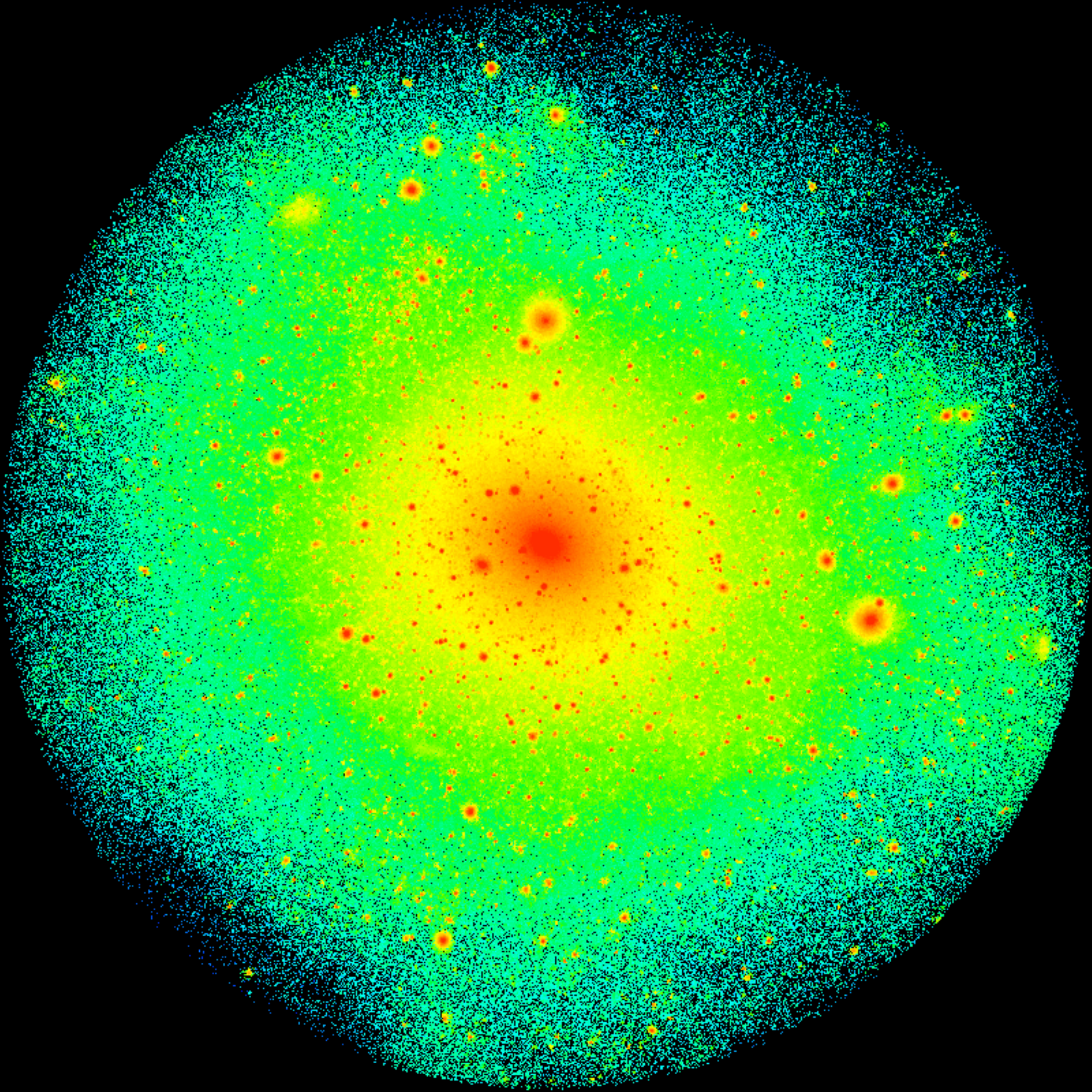}
\includegraphics[scale=0.18]{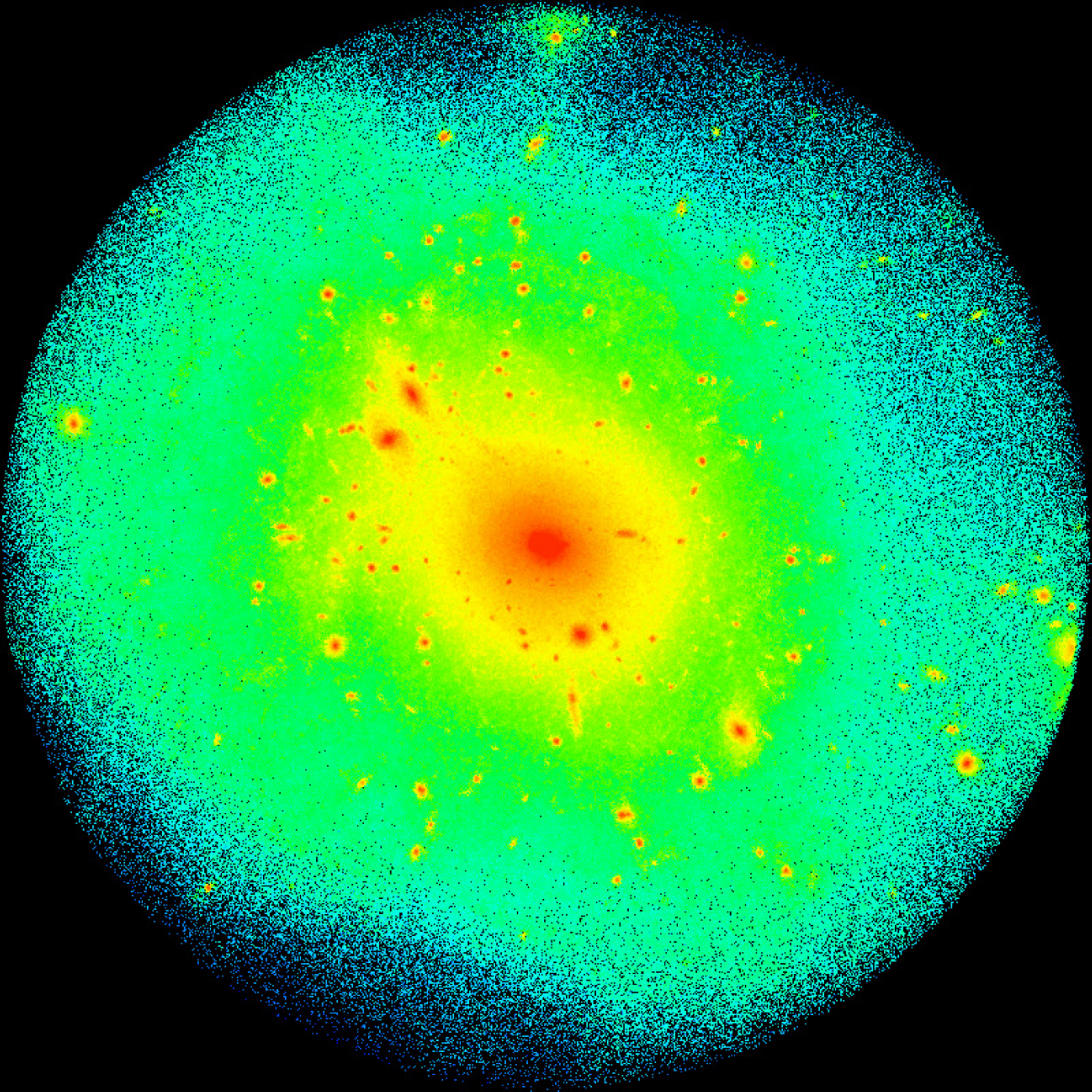}
\caption{Density maps of the refined Milky Way haloes at redshift zero ($R = r_{200}$). From left to right: CDM, WDM 2keV and the most extreme mixed dark matter simulation f20 - 0.1keV.}
\label{galaxymaps}
\end{figure*}

\begin{table}
\caption{Details of the simulations. $R_{200}$ and $N_{200}$ are measured with respect to 200 times the critical density.}
\label{sims}
\begin{center}
\begin{tabular}{lccccc}
\hline
\hline 
Label & m$_{\text{WDM}}$ & $\bar{f}_{\rm W}$ & $\sigma_{\text{th}}^{z=99}$ & $R_{200}$ & $N_{200}$ \\ 
& [keV] & [\%] & [km s$^{-1}$] & [kpc] & [$\times 10^6$] \\
\hline\hline
CDM & - & 0 & 0 & 367 & 9.88 \\
\hline
f05-0.05keV & 0.05 & 5 & 82.1 & 362 & 8.98 \\
f05-0.1keV & 0.1 & 10 & 32.6 & 368 & 9.46 \\
f20-0.1keV & 0.1 & 20 & 51.7 & 365 & 8.49\\ 
f20-0.3keV & 0.3 & 20 & 11.9 & 368 & 9.20  \\ 
f50-0.3keV & 0.3 & 50 & 16.2 & 367 & 9.01\\
f80-1keV & 1.0 & 80 & 3.8 & 368 & 8.69\\
\hline
WDM & 2.0 & 100 & 1.6 & 367 & 9.04 \\
\hline\hline
\end{tabular}
\end{center}
\end{table}
The amount of cold and warm particles in the simulation is the same ($2 \times 256^3$) and their masses are set by the fraction $\bar{f}_{\rm C,W}$, i.e. they differ for $\bar{f}_{\rm C,W} \neq 0.5$. In order to avoid mass segregation as much as possible, the two particle species have different softening lengths, proportional to the fraction $\bar{f}_{\rm C,W}$ \cite{Zempetal2008},
\begin{equation}
\epsilon_{\text{warm}} = \epsilon_0 \cdot \bar{f}_{\rm W}^{1/3}, \qquad \epsilon_{\text{cold}} = \epsilon_0 \cdot  \bar{f}_{\rm C}^{1/3},
\end{equation}
where $\epsilon_0$ is chosen to be 1/50 of the mean inter-particle separation. Different gravitational softening lengths can in principle lead to a non conservation of momentum. However, {\sc pkdgrav} solves this problem by a symmetrization of forces at all times.

The internal structure of dark matter haloes is influenced by thermal motion of warm dark matter particles \cite{maccio2012}. We add a thermal velocity component to all warm particles, computed as a function of the particle mass according to Eq. \eqref{velocitydispersion} \cite{Bodeetal2001}. We follow the procedure described in Ref. \cite{maccio2012}, i.e. creating a three dimensional set of Gaussian randomly distributed velocities corresponding to the three velocity components of each particle and adding them to the Zeldovich velocities at initial redshift, $z = 99$. The variance of the distribution is dependent on the cosmology and is calculated via $\sigma(z) = 3.571 \times v(z)$.
  

\section{The Inner Halo Structure}
\label{internalstructure}
An important characteristic of any non-standard cosmology is the possibility to obtain cored density profiles. In the following we present the measured densities as well as the pseudo phase space densities of the different Milky Way halo realizations. To illustrate the effect of cosmologies with non-zero free streaming, our refined Milky Way halo is shown in Fig. \ref{galaxymaps} for three different cosmologies: CDM, WDM 2keV and our most extreme mixed dark matter model, i.e. f20 - 0.1 keV. The effect on the substructure number density is clearly visible.
\subsection{Density Profiles}
\begin{figure}
\centering
\includegraphics[scale=0.5]{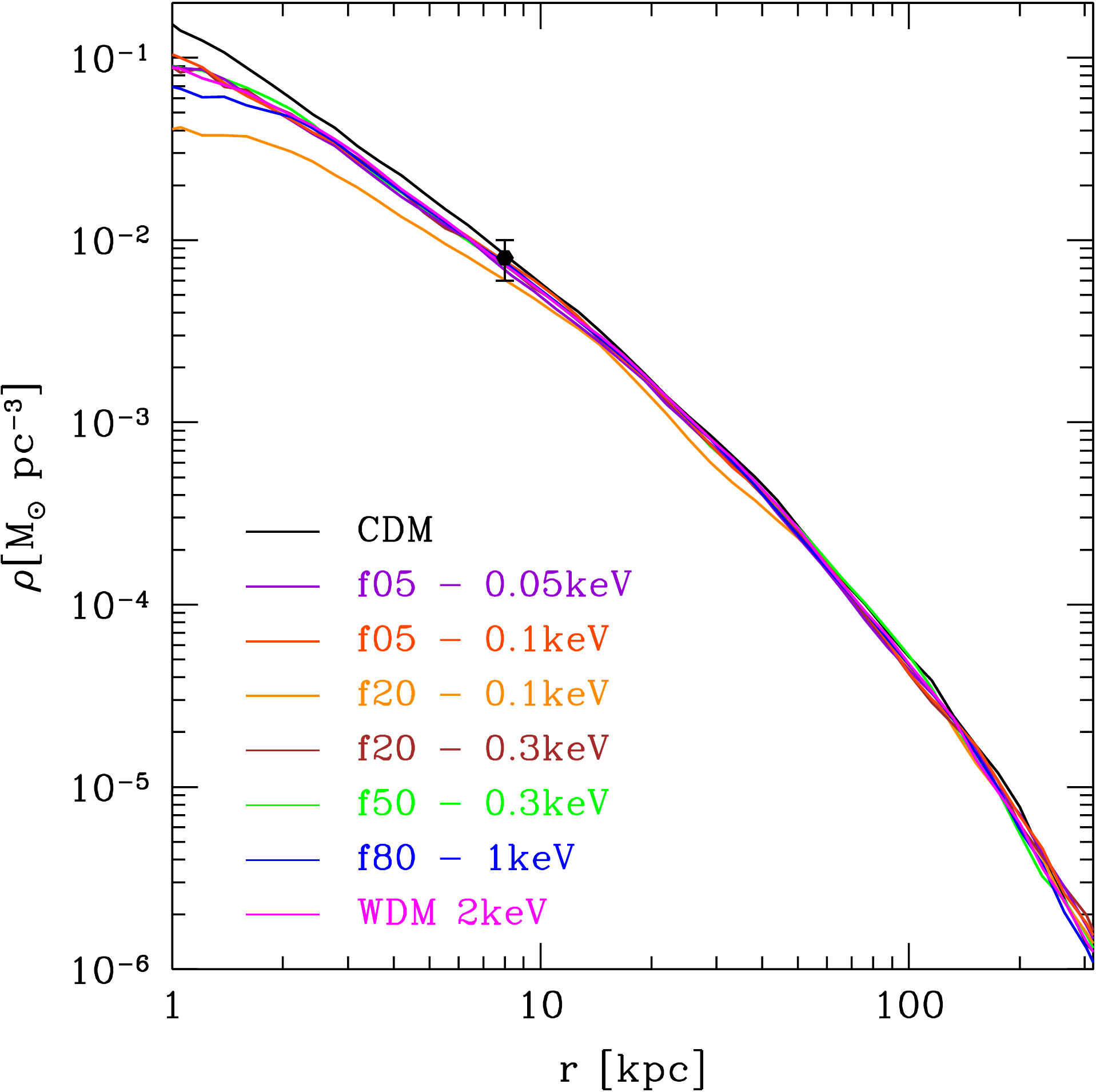}
\caption{The spherically averaged density profiles of the Milky Way sized halo in all eight cosmologies at redshift $z=0$. The data point represents the most recent value for the local dark matter density \cite{BovyTremaine2012}, i.e. $\rho_{\odot} = 0.008\, \pm$ 0.002 M$_{\odot}$ pc$^{-3}$.}
\label{densityprofiles}
\end{figure}

All density profiles of the simulations at redshift $z=0$ listed in Table \ref{sims} are shown in Fig. \ref{densityprofiles}. While the CDM halo shows the usual cuspy behavior in the central region\footnote{For completeness, we fitted the CDM run with three different fitting functions, of which the Einasto profile provides the highest accuracy. The best fit is skipped in Fig. \ref{densityprofiles} for the purpose of simplicity.}, a flattening becomes noticeable in all cosmologies containing a certain fraction of warm dark matter. The central densities in the innermost regions are damped by at least 20 per cent, even if the warm component is subdominant. In the f20 - 0.1 keV model a strong flattening is already evident at $r\sim 10$ kpc.

In principle all mixed dark matter models can be brought into a hierarchical structure, ordered after its free streaming length. Considering a thermal relic with a Fermi-Dirac distribution and from the requirement to provide a correct dark matter abundance, it can be shown that the effective free streaming scale $\lambda$ is proportional to the fraction of warm dark matter times the average thermal velocity of the warm component \cite{Oleg}:
\begin{equation}\label{freestreaming}
\lambda \propto \Big(\frac{\bar{f}_{\rm W}}{m_{\text{WDM}}}\Big)^{4/3}.
\end{equation}
The most extreme simulation according to Eq. \eqref{freestreaming}, i.e. f20 - 0.1 keV, indeed shows the largest flattening of the inner density profile. However, the density profiles of all other cosmologies are not uniquely determined by Eq. \eqref{freestreaming}, which should only be interpreted as a proxy for an effective free streaming of the combined, i.e. cold plus warm, mixed dark matter fluid.\footnote{This hierarchy is more apparent in the mass function, since the number density of collapsed objects is strongly dependent on the cutoff scale of the matter power spectrum, which in turn is a result of free streaming effects in the early universe.}

\begin{figure}
\centering
\includegraphics[scale=0.5]{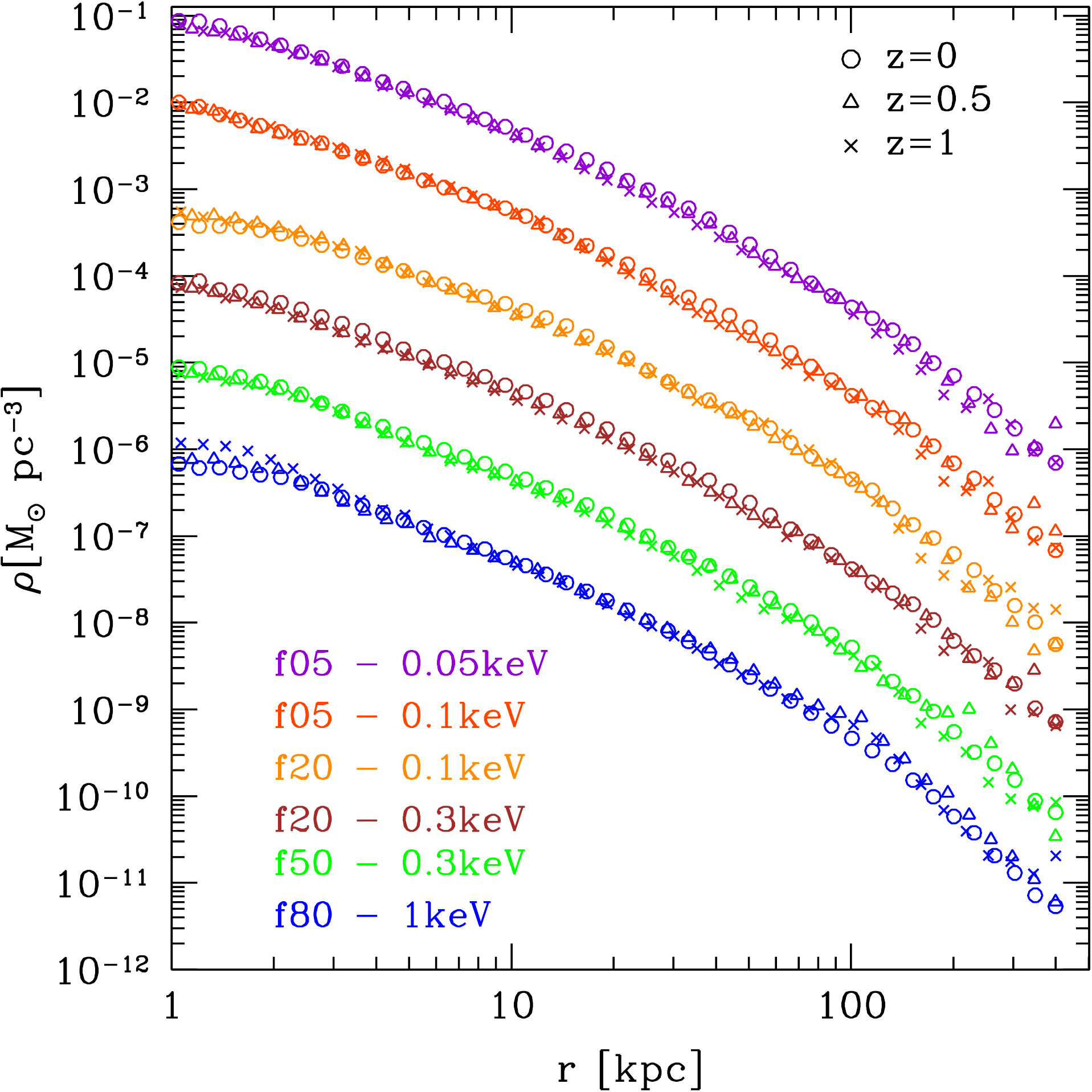}
\caption{Redshift evolution of the density profiles of all six mixed dark matter cosmologies. For a better understanding, only the density profile on top (f05 - 0.05keV) is plotted in physical units whereas the subsequent profiles are each shifted by an additional factor of ten.}
\label{densityevolution}
\end{figure}
\begin{figure}
\centering
\includegraphics[scale=0.5]{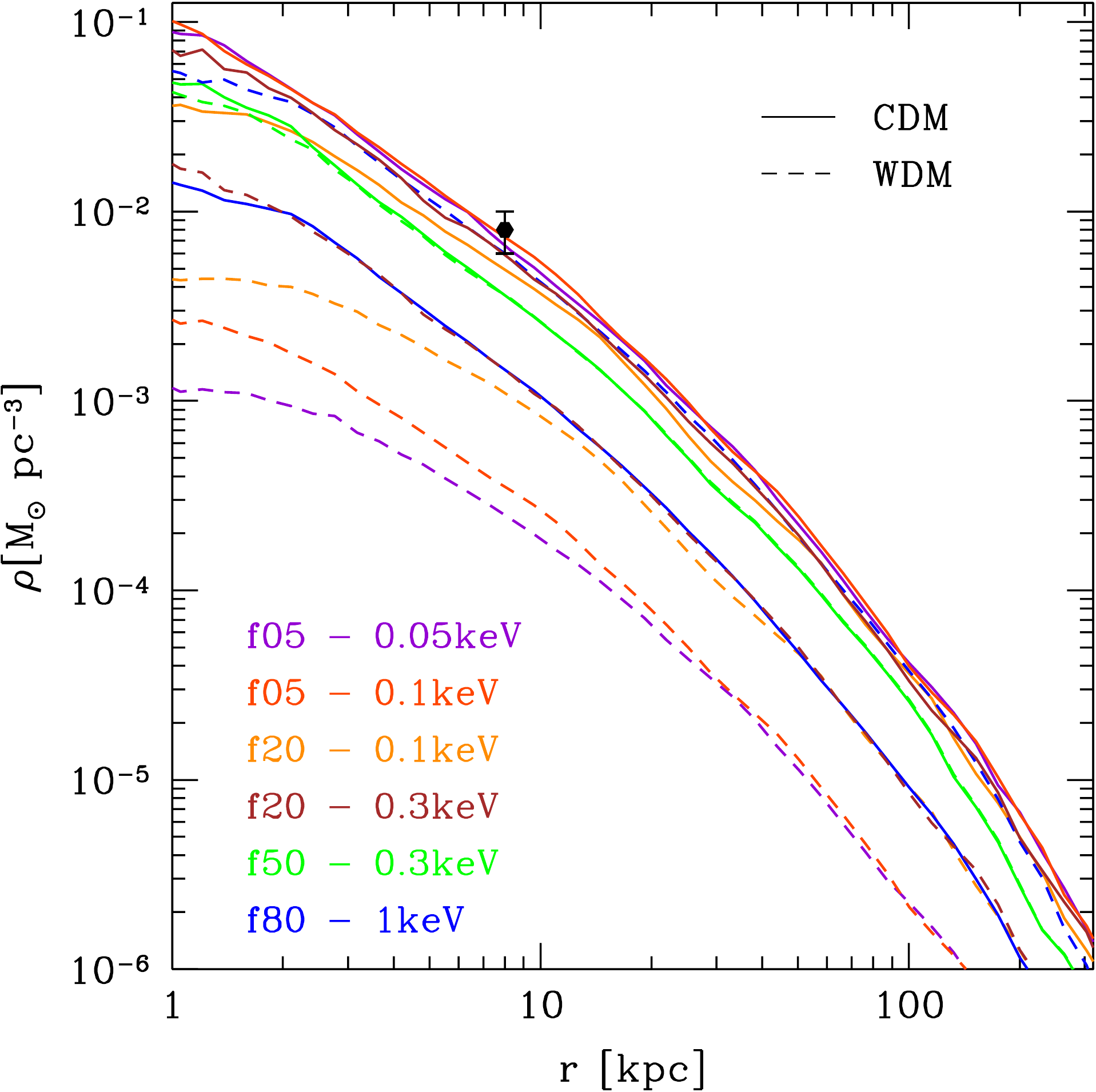}
\caption{Density profiles of the mixed dark matter models, subdivided into the cold (solid) and the warm (dashed) components. For comparison, the most recent value for the local total dark matter density \cite{BovyTremaine2012} is plotted.}
\label{densitycomponents}
\end{figure}
Fig. \ref{densityevolution} shows the time evolution of the density profiles from redshift $z=1$ until today. In all cosmologies, the halo has already fully virialized at redshift $z=1$ and also the flattening of the inner density profile does not substantially evolve until $z=0$. The profiles only show little evolution in the outer parts between redshifts $z=1$ and $z=0$, mostly due to infall of larger subhaloes.
Fig. \ref{densitycomponents} shows the density profiles of each component. Fixing the amount of warm dark matter ($\bar{f}_{\rm W}$), and at the same time varying its mass, creates large differences in the warm component. It differs by almost a factor of two at $r \approx 1$ kpc, while there are almost no differences the cold component. The outer part ($r \gtrsim 40$ kpc) of the halo on the other hand is independent of the WDM mass and is mostly determined by the fraction $\bar{f}_{\rm W}$. This is even true in the case for a warm dark matter dominated MDM model (see f80 - 1keV in Fig.\ref{densitycomponents}). For comparison, Figs. \ref{densityprofiles} \& \ref{densitycomponents} show a estimated local dark matter density \cite{BovyTremaine2012}. The total dark matter density profiles of all models lie in the same range as the measurements of the local density.
\subsection{Pseudo Phase Space Density Profiles}
\begin{figure}
\centering
\includegraphics[scale=0.5]{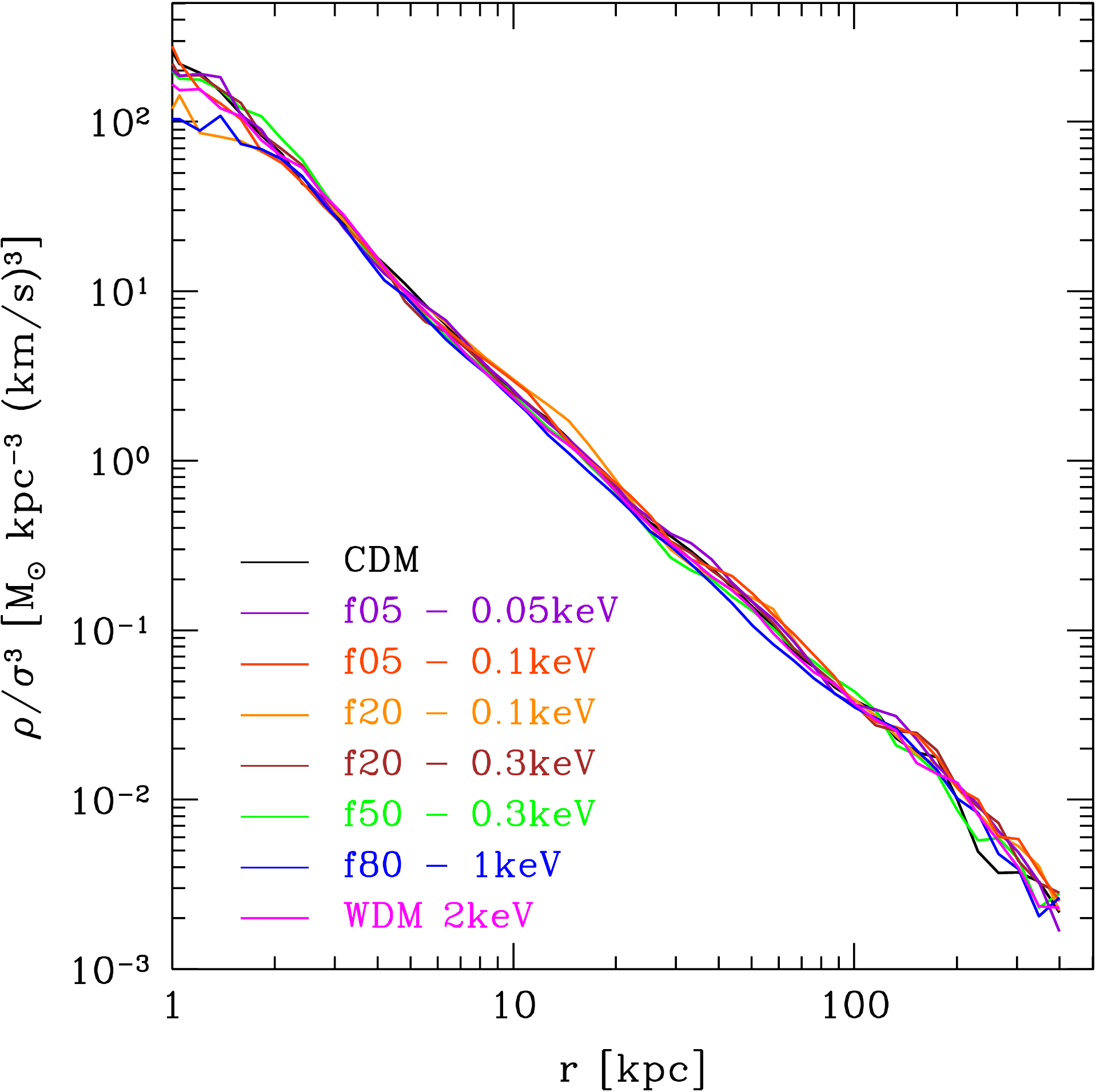}
\caption{Pseudo phase space densities of all mixed dark matter cosmologies, as well as the pure CDM and WDM simulations.}
\label{phasespacedensity}
\end{figure}

As pointed out by earlier works (e.g. \cite{TG79,maccio2012} and references therein), the formation of a core is related to the presence of a maximum in the phase space density distribution. It is useful to introduce the pseudo phase space density \cite{TaylorNavarro2001}, $Q \equiv \rho / \sigma^3$, where $\rho$ is the dark matter density and $\sigma$ is the velocity dispersion. When $Q$ is measured in CDM simulations, it usually follows a power $Q(r)\propto r^{-\alpha}$, with $1.8 \lesssim \alpha \lesssim 1.95$. Fig. \ref{phasespacedensity} shows the pseudo phase space densitiy $Q$ for all the mixed models as well as the pure CDM and the pure WDM case. Some of the models deviate from the power law behaviour ($\alpha \approx 1.9$) only in the very central region of the halo. \cite{maccio2012} showed that a core comparable with observations is only created in a pure warm dark matter scenario with a mass particle candidate, which is in conflict with recent Lyman-$\alpha$ forest data. Since we only ran a limited set of C+WDM simulations, we cannot tackle the question of whether or not such mixed dark matter models can be used to solve the cusp-core problem.

\begin{figure}
\centering
\includegraphics[scale=0.5]{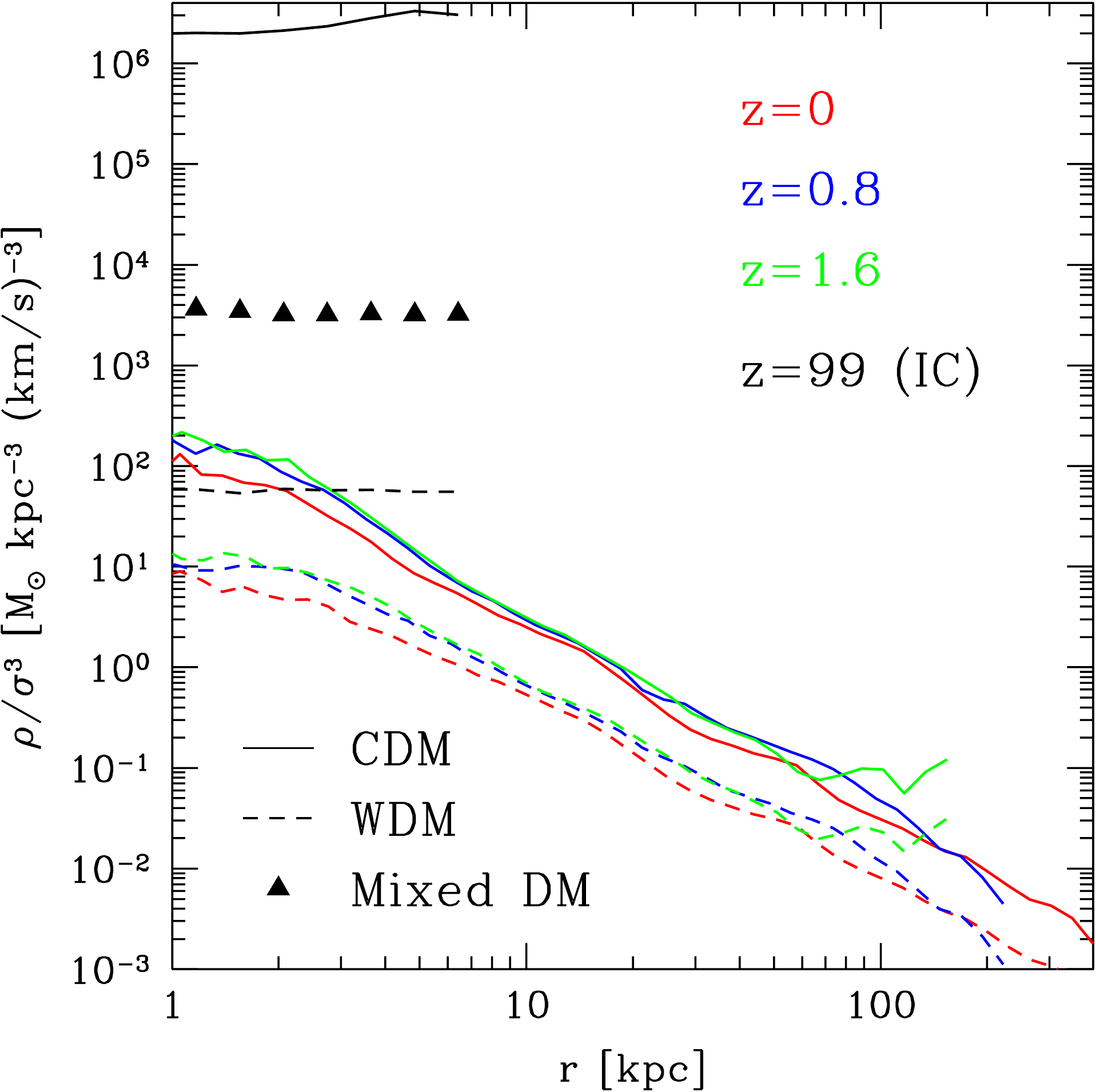}
\caption{Redshift evolution of the phase space densities for each the cold and the warm component. Shown is the example of our most extreme model, i.e. f20 - 0.1 keV Mixed Dark Matter.}
\label{psdevolution}
\end{figure}

Fig. \ref{psdevolution} shows the time evolution of the pseudo phase space density component-by-component for our most extreme model, i.e. f20 - 0.1 keV. The black solid line represents $Q$ of the cold component in the initial conditions ($z=99$), the black dashed line the corresponding warm component. The black triangles on the other hand refer to the total pseudo phase space density in the initial conditions. It is evident that the warm component shows little evolution, whereas the cold component drops by almost five orders of magnitude.

\subsection{Local Dark Matter Densities}
\begin{figure}
\centering
\includegraphics[scale=0.5]{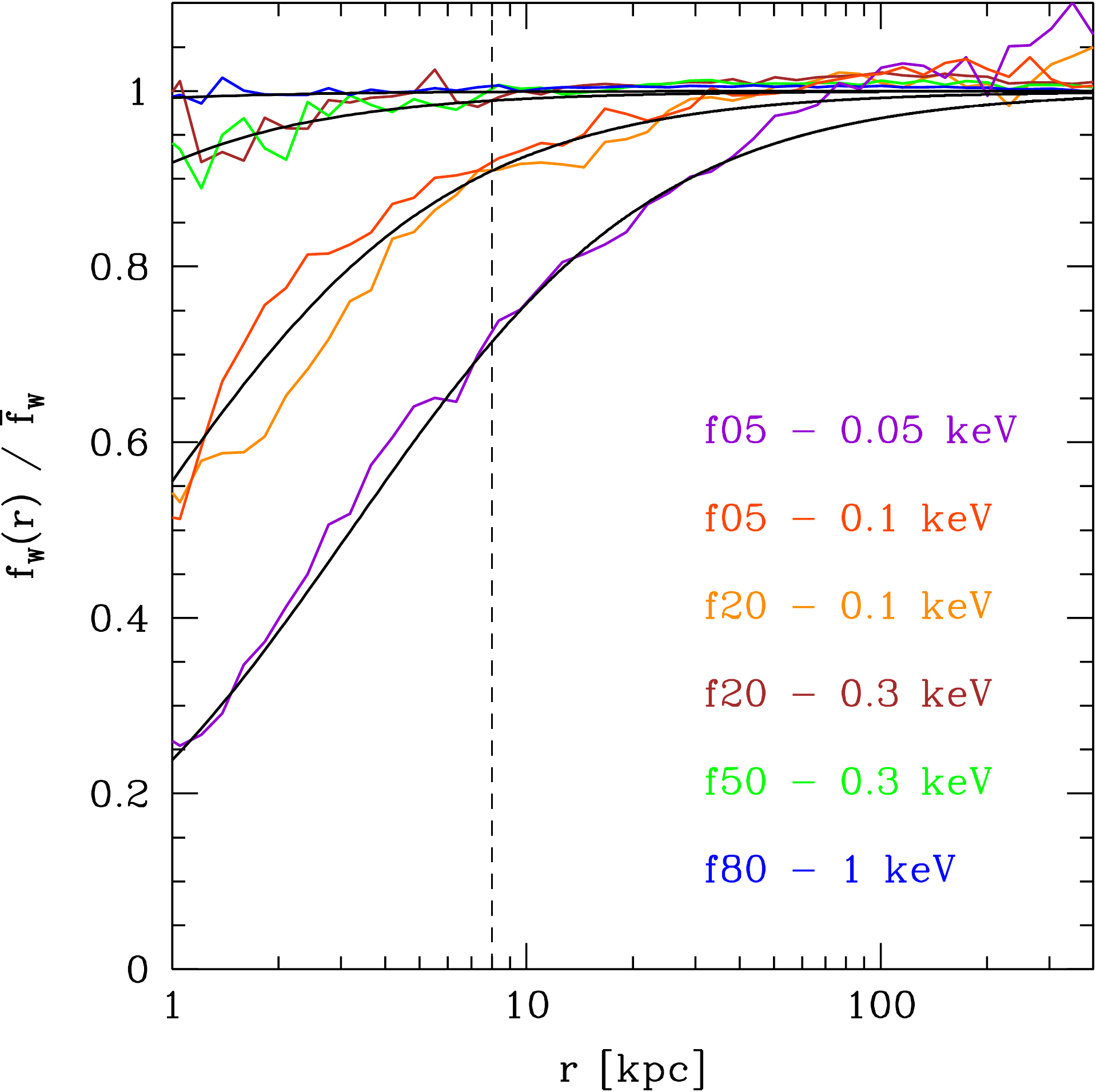}
\caption{Local warm dark matter contribution normalized to the universal WDM fraction. The solid black lines show the fitting function of Eq. \eqref{localdensityfit}. The dashed line indicates the position of the sun in the Milky Way ($r_{\odot} \approx$ 8 kpc).}
\label{densityratio}
\end{figure}

Unlike the C+WDM simulations in \cite{Boyarskyetal2009a,MaccioMixed}, where every simulation particle represents the mixture of both species,
our numerical approach allows us to investigate the composition of the local dark matter in a general C+WDM scenario. Let us define the local density ratio of either the cold or the warm component,
\begin{equation}\label{fi}
f_{i}(r) \equiv \frac{\rho_{i}(r)}{\rho_{\rm C}(r)+\rho_{\rm W}(r)} \,\,\;,\,\,\, i={\rm C,W}.
\end{equation}
Fig. \ref{densityratio} shows the contribution of WDM to the local dark matter density as a function of galactocentric
distance, normalized to the universal fraction of warm dark matter, i.e. $(f_{\rm W}$ / $\bar{f}_{\rm W})$.
This quantity is found to be independent on $\bar{f}_{\rm W}$, but to dependent strongly on the warm dark matter mass. The dashed vertical line in Fig. \ref{densityratio} further shows the position of the sun in the Milky Way and the black solid lines refer to an empirical fitting formula for the normalized local WDM fraction, given by
\begin{equation}\label{localdensityfit}
\mathcal{F}(m,r) = \Big[1 + \xi \cdot \Big(\frac{\text{keV}}{m_{\text{WDM}}}\Big)^2 \Big(\frac{\rm kpc}{r}\Big) \Big]^{-1}.
\end{equation}
Using the Levenberg-Marquart method, we found that $\xi = 0.008$ fits all C+WDM models simultaneously with high accuracy. The deviations at small radii are probably due to higher order effects.

We want to emphasize that Eq. \eqref{localdensityfit} is explicitly valid only for a Milky Way size dark matter halo, since we expect the radial evolution of the warm/cold ratio to be dependent on the halo mass. In a pure warm dark matter cosmology, assuming isothermal density profiles, there is a dependence between halo mass and the expected core radius due to free streaming \cite{HoganDalcanton2000,maccio2012}, $r_c \propto \sigma^{-1/2} \propto M_{\rm halo}^{-1/6}$. We therefore expect the local density ratio at a fixed radius to be closer to the universal value when moving from Milky Way to cluster sized dark matter haloes. Additional simulations of dwarf ($10^{10}$ M$_{\odot}$)
and cluster ($10^{14}$ M$_{\odot}$) haloes are planned and will allow to quantify
the halo mass dependence of these density ratios in detail.
\section{Implications for Dark Matter detection experiments}
\label{detection}

Astroparticle searches for dark matter particles can be broadly divided into {\it direct} and {\it indirect} searches (see \cite{Bergstrom:2000pn, Munoz:2003gx,Bertone:2004pz,Bertone:2010at,Feng:2010g} for recent reviews). In this section, we discuss the implications of our results for both detection strategies.

Direct dark matter searches aim to detect the recoil energy of detector nuclei struck by DM particles, assumed to be in the form of Weakly Interacting Massive Particles (WIMPs), a prototypical example of CDM candidates. The expected rate of events $R$ is proportional to the product of the scattering cross section of WIMPs off baryons $\sigma_{cn}$ times the flux of CDM particles through the detector, which in turn is proportional to the local dark matter density, therefore
\begin{equation}
R \propto \sigma_{cn} \rho_{\rm C,\odot}.
\end{equation}
In the literature, it is often assumed that $\rho_{\rm C,\odot} = \rho_{\rm DM,\odot}$, i.e. that CDM completely saturates the {\it measured} DM density $\rho_{DM,\odot} \approx 0.3 $ GeV cm$^{-3}$ (see \cite{Catena:2009mf, Pato:2010yq,Salucci:2010qr,Iocco:2011jz,Garbarietal2011,BovyTremaine2012} or a discussion of uncertainties on this quantity). This corresponds to the assumption that  all the dark matter in the solar neighborhood is in the form of CDM particles, $f_{\rm C,\odot} = 1$. As we have seen, however, dark matter could in principle be made of a mixture of particles, in which case this assumption is explicitly violated. 

This is particularly important at a time when the Large Hadron Collider (LHC) is probing the parameter space of some of the most popular candidates for dark matter, such as the supersymmetric neutralino. In case of discovery, the question of whether the newly discovered particles compose all of the DM in the Universe can probably be  answered only through a combined analysis that also includes direct and/or indirect detection experiments \cite{Bertone:2010at}.   

In order to combine LHC and direct detection data one can follow two strategies \cite{Bertone:2010rv}. The first is to identify CDM with DM locally {\it and} on average in the Universe. Since the aim of the analysis is to prove that the new particles compose all of the DM in the Universe, this analysis is clearly not satisfactory and it can at best provide a {\it consistency check} for a DM interpretation of LHC data.

To go beyond these simplistic assumptions, one can make the much less stringent {\it ansatz} $f_{\rm C,\odot} = \bar{f_{\rm C}}$  \cite{Bertone:2010rv}, which corresponds to assume that the fraction of CDM is the same locally as on average in the Universe. The rate of events can therefore be written as
\begin{equation}
R \propto \sigma_{cn} \rho_{\rm C,\odot} = \sigma_{cn} \rho_{\odot} f_{\rm C,\odot} = \sigma_{cn} \rho_{\odot} \bar{f}_{\rm C} 
\end{equation}
Since this assumption proves to be very effective in breaking the degeneracies in the parameter estimations of DM properties for combined LHC/direct detection searches, it is worth understanding the limits of its applicability to the problem at hand in light of the results discussed in the previous section.

In the limit where the two (or more) components of dark matter are perfectly cold, one does not expect any deviations from the scaling ansatz above, since the two components are perfectly mixed in galactic halos down to the smallest scales. The results of our simulations allow us to precisely identify the regime where the $f_{\rm C}(r) / \bar{f}_{\rm C}$ starts to deviate from unity, i.e. when the scaling ansatz fails. Fig. \ref{densityratio} accurately provides this information, since it shows the behavior of this ratio as a function of the mass of the warm component and the distance from the galactic center. 

In Fig. \ref{densityratio} we observe deviations from perfect scaling at the solar radius by up to 30\% for our models, although even larger deviations are possible (see Eq. \eqref{localdensityfit}). Therefore, using the scaling ansatz introduces a significant systematic uncertainty and one should use Eq. \eqref{localdensityfit} for a better estimate of the local densities. We also stress that the presence of a WDM component always leads to the inequality $f_{\rm C} \geq \bar{f}_{\rm C}$.

As for indirect detection, which is based on the observation of the annihilation or decay products of DM particles, we find as expected opposite effects on strategies for WDM and CDM searches. As we have seen, in fact, the joint evolution of these two components tends to increase the relative contribution of CDM over WDM inside virialized structures. The specific enhancement (reduction) of gamma-ray (X-ray) fluxes expected from the annihilation (decay) of CDM (WDM) particles depends on the specific observational target and on the properties of the WDM component, and it is outside the aim of this paper. We limit ourselves here to notice that since all annihilation fluxes scale with the square of the number density of CDM particles, the predicted annihilation radiation depends strongly on the specific behaviour of $f_{\rm C}(r)$ in the virtualized halo under consideration.

\section{Summary and Conclusion}
We have performed a set of high resolution simulations of a Milky Way sized halo in different C+WDM cosmologies, varying the fraction and the mass of the warm component. For each run we therefore set up initial conditions, consisting of two distinct sets of dark matter particles, each species initialised with the corresponding modified matter power spectrum. This procedure enables us to track the density profiles of each component and to test the scaling ansatz, i.e. how the density ratio of the cold and warm components changes as a function of distance to the galactic center. We find:
\begin{itemize}
  \item The transfer functions of the cold and the warm component in a general C+WDM cosmology are strongly correlated. Unlike in pure WDM, where $T(k)$ rapidly drops to zero after the cutoff, the presence of a cold component also stabilises the warm component and it approaches a constant plateau at high $k$. This leads to non negligible power in the warm dark matter component even at the smallest scales.
 \vspace{5mm}
  \item Due to thermal velocities of the WDM particles, all mixed dark matter scenarios in this work show significantly reduced total dark matter densities in the halo center. The density profile of our warmest model, i.e. the f20 - 0.1 keV, shows a reduced density already at large radii (8 kpc) and a possible hint of a flat core in the very inner region ($\sim$ 1 kpc).
  \vspace{5mm}
  \item The assumption that the fraction of any component in a mixed dark matter model is the same locally as on average in the Universe, is violated in some of the scenarios. The presence of a warm dark matter component leads to the inequality $f_{\rm W} \leq \bar{f}_{\rm W}$ ($f_{\rm C} \geq \bar{f_{\rm C}}$).
  \vspace{5mm}
  \item The normalised WDM fraction ($f_{\rm W}(r)$ / $\bar{f}_{\rm W}$) is proportional to the squared mass of the WDM particle and it is practically independent on $\bar{f}_{\rm W}$. This is surprising, since $m_{\rm WDM}$ is responsible for the cutoff of the linear C+WDM power spectrum, whereas the fraction determines the plateau at small scales.
   \vspace{5mm}
   \item In absence of further information on the nature of any dark matter component, our results on the local density introduce systematic uncertainties in the data analysis of a combined, i.e. accelerator plus direct detection, dark matter search.
   \vspace{5mm}
   \item For the interpretation of dark matter search data in generic mixed dark matter cosmologies one should use Eq. \eqref{localdensityfit} instead of the scaling ansatz, i.e. $f_{\rm C/W, \odot} = \bar{f}_{\rm C/W}$.
   \end{itemize}

\acknowledgments  
It is a pleasure to thank Oleg Ruchayskiy and Alexey Boyarsky for stimulating discussions as well as Marcel Zemp for useful comments in the final stages of this work. GB acknowledges the support of the European Research Council through the ERC Starting Grant {\it WIMPs Kairos}. Finally, the authors would like to thank the referee for making some very useful comments which have helped to improve the paper. Numerical simulations were performed on the Theo and the PanStarrs2 clusters at the Max-Planck-Institut f\"ur Astronomie at the Rechenzentrum in Garching, the zBox3 cluster at the Institute for Theoretical Physics in Zurich and on the Rosa cluster at the CSCS in Lugano. This work was supported by the SNF.

\end{document}